\documentclass[12pt,tightenlines,eqsecnum,floats,aps,amsmath,amssymb,nofootinbib,prd,shownopacs]{revtex4}
\usepackage{amssymb}
\usepackage{graphicx}
\usepackage{amsmath,amssymb,amsfonts,amsthm}

\def\be{\begin{equation}}
\def\ee{\end{equation}}
\def\ba{\begin{eqnarray}}
\def\ea{\end{eqnarray}}

\def\H{{\cal H}}

\def\u(1){{\rm u(1)}}

\def\lp{{\ell}_{\rm Pl}}

\def\b{$\bullet\,\,$}

\begin{document}


\title{Introduction to Loop Quantum Gravity}
\author{Abhay\ Ashtekar}
\affiliation{Institute for Gravitational Physics and Geometry,\\
Physics Department, 104 Davey, Penn State, University Park, PA
16802, USA}


\begin{abstract}

This article is based on the opening lecture at the third quantum
geometry and quantum gravity school sponsored by the European
Science Foundation and held at Zakopane, Poland in March 2011. The
goal of the lecture was to present a broad perspective on loop
quantum gravity for young researchers. The first part is addressed
to beginning students and the second to young researchers who are
already working in quantum gravity.

\end{abstract}

 \maketitle

\bigskip
\indent\emph{Pacs {04.60Pp, 04.60.Ds, 04.60.Nc, 03.65.Sq}}



\section{Introduction}
\label{s1}

This section, addressed to beginning researchers, is divided into
two parts. The first provides a broad historical perspective and the
second illustrates key physical and conceptual problems of quantum
gravity. Researchers who are already quite familiar with quantum
gravity can/should go directly to section \ref{s2}; there will be no
loss of continuity.

\subsection{Development of Quantum Gravity: A Bird's Eye View}
\label{s1.1}

The necessity of a quantum theory of gravity was pointed out by
Einstein already in a 1916 paper in the Preussische Akademie
Sitzungsberichte. He wrote:

\begin{itemize}

\item \textsl{Nevertheless, due to the inneratomic movement
    of electrons, atoms would have to radiate not only
    electromagnetic but also gravitational energy, if only
    in tiny amounts. As this is hardly true in Nature, it
    appears that quantum theory would have to modify not
    only Maxwellian electrodynamics but also the new theory
    of gravitation.} \b

\end {itemize}
Papers on the subject began to appear in the 1930s most notably
by Bronstein, Rosenfeld and Pauli. However, detailed work began
only in the sixties. The general developments since then loosely
represent four stages, each spanning roughly a decade and a
half. In this section, I will present a sketch these
developments.

First, there was the beginning: exploration.  The goal was to do
unto gravity as one would do unto any other physical field
\cite{cji1}.%
\footnote{Since this introduction is addressed to
non-experts, I will generally refer to books and review
articles which summarize the state of the art at various
stages of development of quantum gravity. References to
original papers can be found in these reviews.}
The electromagnetic field had been successfully quantized
using two approaches: canonical and covariant. In the
canonical approach, electric and magnetic fields obeying
Heisenberg's uncertainty principle are at the forefront,
and quantum states naturally arise as gauge-invariant
functionals of the vector potential on a spatial
three-slice.  In the covariant approach on the on the other
hand, one first isolates and then quantizes the two
radiative modes of the Maxwell field in space-time, without
carrying out a (3+1)-decomposition, and the quantum states
naturally arise as elements of the Fock space of photons.
Attempts were made to extend these techniques to general
relativity.  In the electromagnetic case the two methods
are completely equivalent.  Only the emphasis changes in
going from one to another.  In the gravitational case,
however, the difference is \emph{profound}.  This is not
accidental.  The reason is deeply rooted in one of the
essential features of general relativity, namely the dual
role of the space-time metric.

To appreciate this point, let us begin with field theories
in Minkowski space-time, say Maxwell's theory to be
specific.  Here, the basic dynamical field is represented
by a tensor field $F_{\mu\nu}$ on Minkowski space.  The
space-time geometry provides the kinematical arena on which
the field propagates. The background, Minkowskian metric
provides  light cones and the notion of causality.  We can
foliate this space-time by a one-parameter family of
space-like three-planes, and analyze how the values of
electric and magnetic fields on one of these surfaces
determine those on any other surface.  The isometries of
the Minkowski metric let us construct physical quantities
such as fluxes of energy, momentum, and angular momentum
carried by electromagnetic waves.

In general relativity, by contrast, there is no background
geometry.  The space-time metric itself is the fundamental
dynamical variable.  On the one hand it is analogous to the
Minkowski metric in Maxwell's theory; it determines space-time
geometry, provides light cones, defines causality, and dictates
the propagation of all physical fields (including itself).  On
the other hand it is the analog of the Newtonian gravitational
potential and therefore the basic dynamical entity of the
theory, similar in this respect to the $F_{\mu\nu}$ of the
Maxwell theory. This dual role of the metric is in effect a
precise statement of the equivalence principle that is at the
heart of general relativity. It is this feature that is largely
responsible for the powerful conceptual economy of general
relativity, its elegance and its aesthetic beauty, its
strangeness in proportion. However, this feature also brings
with it a host of problems.  We see already in the classical
theory several manifestations of these difficulties. It is
because there is no background geometry, for example, that it is
so difficult to analyze singularities of the theory and to
define the energy and momentum carried by gravitational waves.
Since there is no a priori space-time, to introduce notions as
basic as causality, time, and evolution, one must first solve
the dynamical equations and \textit{construct} a space-time. As
an extreme example, consider black holes, whose traditional
definition requires the knowledge of the causal structure of the
entire space-time. To find if the given initial conditions lead
to the formation of a black hole, one must first obtain their
maximal evolution and, using the causal structure determined by
that solution, ask if the causal past $J^-(\mathcal{I}^+)$ of
its future infinity $\mathcal{I}^+$ is the entire space-time. If
not, space-time contains a black hole and the future boundary of
$J^(\mathcal{I}^+)$ within that space-time is its event horizon.
Thus, because there is no longer a clean separation between the
kinematical arena and dynamics, in the classical theory
substantial care and effort is needed even in the formulation of
basic physical questions.

In quantum theory the problems become significantly more
serious. To see this, recall first that, because of the
uncertainty principle, already in non-relativistic quantum
mechanics, particles do not have well-defined trajectories;
time-evolution only produces a probability amplitude,
$\Psi(x,t)$, rather than a specific trajectory, $x(t)$.
Similarly, in quantum gravity, even after evolving an initial
state, one would not be left with a specific space-time. In the
absence of a space-time geometry, how is one to introduce even
habitual physical notions such as causality, time, scattering
states, and black holes?

The canonical and the covariant approaches adopted dramatically
different attitudes to face these problems. In the canonical
approach, one notices that, in spite of the conceptual
difficulties mentioned above, the Hamiltonian formulation of
general relativity is well-defined and attempts to use it as a
stepping stone to quantization. The fundamental canonical
commutation relations are to lead us to the basic uncertainty
principle.  The motion generated by the Hamiltonian is to be
thought of as time evolution.  The fact that certain operators
on the fixed (`spatial') three-manifold commute is supposed to
capture the appropriate notion of causality. The emphasis is on
preserving the geometrical character of general relativity, on
retaining the compelling fusion of gravity and geometry that
Einstein created.  In the first stage of the program, completed
in the early 1960s, the Hamiltonian formulation of the classical
theory was worked out in detail by Dirac, Bergmann, Arnowitt,
Deser and Misner and others \cite{adm,komar,pbak,agrev,kk1}. The
basic canonical variable was the 3-metric on a spatial slice.
The ten Einstein's equations naturally decompose into two sets:
four constraints on the metric and its conjugate momentum
(analogous to the equation ${\rm Div} \vec{E} = 0$ of
electrodynamics) and six evolution equations. Thus, in the
Hamiltonian formulation, general relativity could be interpreted
as the dynamical theory of 3-geometries. Wheeler therefore
baptized it \emph{geometrodynamics} \cite{jw1,jw2}.

In the second stage, this framework was used as a point of
departure for quantum theory by Bergmann, Komar, Wheeler DeWitt
and others. The basic equations of the quantum theory were
written down and several important questions were addressed
\cite{jw2,kk1}. Wheeler also launched an ambitious program in
which the internal quantum numbers of elementary particles were
to arise from non-trivial, microscopic topological
configurations and particle physics was to be recast as
`chemistry of geometry'. However, most of the work in quantum
geometrodynamics continued to remain formal; indeed, even today
the field theoretic difficulties associated with the presence of
an \emph{infinite number of degrees of freedom} in the Wheeler
DeWitt equation remain unresolved. Furthermore, even at the
formal level, is has been difficult to solve the quantum
Einstein's equations. Therefore, after an initial burst of
activity, the quantum geometrodynamics program became stagnant.
Interesting results have been obtained by Misner, Wheeler,
DeWitt and others in the limited context of quantum cosmology
where one freezes all but a finite number of degrees of freedom.
However, even in this special case, the initial singularity
could not be resolved without additional `external' inputs into
the theory, such as the use of matter violating energy
conditions. Sociologically, the program faced another
limitation: concepts and techniques which had been so successful
in quantum electrodynamics appeared to play no role here. In
particular, in quantum geometrodynamics, it is hard to see how
gravitons are to emerge, how scattering matrices are to be
computed, how Feynman diagrams are to dictate dynamics and
virtual processes are to give radiative corrections. To use a
well-known phrase \cite{weinberg}, the emphasis on geometry in
the canonical program ``drove a wedge between general relativity
and the theory of elementary particles."

In the covariant%
\footnote{In the context of quantum gravity, the term
`covariant' is somewhat misleading because the introduction
of a background metric violates diffeomorphism covariance.
It is used mainly to emphasize that this approach does not
involve a 3+1 decomposition of space-time.}
approach \cite{agrev,bsd,md} the emphasis is just the
opposite. Field-theoretic techniques are put at the
forefront. The first step in this program is to split the
space-time metric $g_{\mu\nu}$ in two parts, $g_{\mu\nu}=
\eta_{\mu\nu} + \sqrt{G}\, h_{\mu\nu}$, where
$\eta_{\mu\nu}$ is to be a background, kinematical metric,
often chosen to be flat, $G$ is Newton's constant, and
$h_{\mu\nu}$, the deviation of the physical metric from the
chosen background, the dynamical field. The two roles of
the metric tensor are now split. The overall attitude is
that this sacrifice of the fusion of gravity and geometry
is a moderate price to pay for ushering-in the powerful
machinery of perturbative quantum field theory. Indeed,
with this splitting most of the conceptual problems
discussed above seem to melt away. Thus, in the transition
to the quantum theory it is only $h_{\mu\nu}$ that is
quantized.  Quanta of this field propagate on the classical
background space-time with metric $\eta_{\mu\nu}$. If the
background is in fact chosen to be flat, one can use the
Casimir operators of the Poincar\'e group and show that the
quanta have spin two and rest mass zero.  These are the
gravitons. The Einstein-Hilbert Lagrangian tells us how
they interact with one another.  Thus, in this program,
quantum general relativity was first reduced to a quantum
field theory in Minkowski space. One could apply to it all
the machinery of perturbation theory that had been so
successful in particle physics.  One now had a definite
program to compute amplitudes for various scattering
processes.  Unruly gravity appeared to be tamed and forced
to fit into the mold created to describe other forces of
Nature. Thus, the covariant quantization program was more
in tune with the mainstream developments in physics at the
time. In 1963 Feynman extended perturbative methods from
quantum electrodynamics to gravity. A few years later
DeWitt carried this analysis to completion by
systematically formulating the Feynman rules for
calculating scattering amplitudes among gravitons and
between gravitons and matter quanta. He showed that the
theory is unitary order by order in the perturbative
expansion. By the early seventies, the covariant approach
had led to several concrete results \cite{bsd}.

Consequently, the second stage of the covariant program
began with great enthusiasm and hope. The motto was: Go
forth, perturb, and expand. The enthusiasm was first
generated by the discovery that Yang-Mills theory coupled
to fermions is renormalizable (if the masses of gauge
particles are generated by a spontaneous
symmetry-breaking mechanism).%
\footnote{In fact DeWitt's quantum gravity work \cite{bsd} played
a seminal role in the initial stages of the extension of
perturbative techniques from Abelian to non-Abelian gauge
theories.}
This led to a successful theory of electroweak
interactions. Particle physics witnessed a renaissance of
quantum field theory. The enthusiasm spilled over to
gravity. Courageous calculations were performed to estimate
radiative corrections.  Unfortunately, however, this
research soon ran into its first road block. The theory was
shown to be non-renormalizable when two loop effects are
taken into account for pure gravity and already at one loop
for gravity coupled with matter \cite{cji2}. To appreciate
the significance of this result, let us return to the
quantum theory of photons and electrons. This theory is
perturbatively renormalizable.  This means that, although
individual terms in the perturbation expansion of a
physical amplitude may diverge due to radiative corrections
involving closed loops of virtual particles, these
infinities are of a specific type; they can be
systematically absorbed in the values of free parameters of
the theory, the fine structure constant and the electron
mass.  Thus, by renormalizing these parameters, individual
terms in the perturbation series can be systematically
rendered finite. In quantum general relativity, such a
systematic procedure is not available; infinities that
arise due to radiative corrections are genuinely
troublesome.  Put differently, quantum theory acquires an
infinite number of undetermined parameters. Although one
can still use it as an effective theory in the low energy
regime, regarded as a fundamental theory, it has no
predictive power at all!

Buoyed, however, by the success of perturbative methods in
electroweak interactions, the particle physics community was
reluctant to give them up in the gravitational case. In the case
of weak interactions, it was known for some time that the
observed low energy phenomena could be explained using Fermi's
simple four-point interaction. The problem was that this Fermi
model led to a non-renormalizable theory. The correct,
renormalizable model of Glashow, Weinberg and Salam agrees  with
Fermi's at low energies but marshals new processes at high
energies which improve the ultraviolet behavior of the theory.
It was therefore natural to hope that the situation would be
similar in quantum gravity. General relativity, in this analogy,
would be similar to Fermi's model. The fact that it is not
renormalizable was taken to mean that it ignores important
processes at high energies which are, however, unimportant at
low energies, i.e., at large distances. Thus, the idea was that
the correct theory of gravity would differ from general
relativity but only at high energies, i.e., near the Planck
regime. With this aim, higher derivative terms were added to the
Einstein-Hilbert Lagrangian. If the relative coupling constants
are chosen judiciously, the resulting theory does in fact have a
better ultraviolet behavior. Stelle, Tomboulis and others showed
that the theory is not only renormalizable but asymptotically
free; it resembles the free theory in the high energy limit.
Thus, the initial hope of `curing' quantum general relativity
was in fact realized. However, it turned out that the
Hamiltonian of this theory is unbounded from below, and
consequently the theory is drastically unstable! In particular,
it violates unitarity; probability fails to be conserved. The
success of the electroweak theory suggested a second line of
attack. In the approaches discussed above, gravity was
considered in isolation. The successful unification of
electromagnetic and weak interactions suggested the possibility
that a consistent theory would result only when gravity is
coupled with suitably chosen matter. The most striking
implementation of this viewpoint occurred in supergravity. Here,
the hope was that the bosonic infinities of the gravitational
field would be canceled by those of suitably chosen fermionic
sources, giving us a renormalizable quantum theory of gravity.
Much effort went into the analysis of the possibility that the
most sophisticated of these theories ---$N = 8$ supergravity---
can be employed as a genuine grand unified
theory.%
\footnote{For a number of years, there was a great deal of
confidence, especially among particle physicists, that
supergravity was on the threshold of providing the complete
quantum gravity theory. For instance, in the centennial
celebration of Einstein's birthday at the Institute of Advanced
Study, Princeton \cite{wolf} ---the proceedings of which were
videotaped and archived for future historians and physicists---
there were two talks on quantum gravity, both devoted to
supergravity. A year later, in his Lucasian Chair inaugural
address Hawking \cite{swh1} suggested that end of theoretical
physics was in sight because $N=8$ supergravity was likely to be
the final theory.}
It turned out that cancelations of infinities do occur. Over the
last five years or so, there has been a resurgence of interest
in this area \cite{bern}. It has now been shown that
supergravity is finite to four loops even though it contains
matter fields coupled to gravity. Furthermore, its Hamiltonian
is manifestly positive and the theory is unitary. However, there
are several arguments suggesting that the theory would not be
renormalizable at seven loops. This is still an open issue but
could be settled in the next 2-3 years.

By and large, the canonical approach was pursued by relativists
and the covariant approach by particle physicists. In the mid
1980s, both approaches received unexpected boosts. These
launched the third phase in the development of quantum gravity.

A group of particle physicists had been studying string theory to
analyze strong interactions from a novel angle. The idea was to
replace point particles by 1-dimensional extended objects
---strings--- and associate particle-like states with various
modes of excitations of the string. Initially there was an
embarrassment: in addition to the spin-1 modes characteristic of
gauge theories, string theory included also a spin-2, massless
excitation. But it was soon realized that this was a blessing in
disguise: the theory automatically incorporated a graviton. In
this sense, gravity was already built into the theory! However,
it was known that the theory had a potential quantum anomaly
which threatened to make it inconsistent. In the mid 1980s,
Greene and Schwarz showed that there is an anomaly cancelation
and perturbative string theory could be consistent in certain
space-time dimensions ---26 for a purely bosonic string and 10
for a superstring \cite{gsw,jpbook}. Since strings were assumed
to live in a flat background space-time, one could apply
perturbative techniques. However, in this reincarnation, the
covariant approach underwent a dramatic revision. Since it is a
theory of extended objects rather than point particles, the
quantum theory has brand new elements; it is no longer a local
quantum field theory. The field theoretic Feynman diagrams are
replaced by world-sheet diagrams. This replacement dramatically
improves the ultraviolet behavior and, although explicit
calculations have been carried out only at 2 or 3 loop order, in
the string theory community it is widely believed that the
perturbation theory is \emph{finite} to all orders; it does not
even have to be renormalized. The theory is also unitary. It has
a single, new fundamental constant ---the string tension--- and,
since various excited modes of the string represent different
particles, there is a built-in principle for unification of all
interactions!%
\footnote{To date, none of the low energy reductions appears to
correspond to the world we actually observe. Nonetheless, string
theory has provided us with a glimpse of an entirely new vista:
the concrete possibility that unification could be brought about
by a tightly woven, non-local theory.}
{}From the viewpoint of local quantum field theories that
particle physicists have used in studying electroweak and strong
interactions, this mathematical structure seems almost magical.
Therefore there has been a long-standing hope in the string
community that this theory would encompass all of fundamental
physics; it would be the `theory of everything'.

Unfortunately, it soon became clear that string perturbation
theory also faces some serious limitations. Perturbative
finiteness would imply that each term in the perturbation
series is ultra-violet finite.%
\footnote{But it does appear that there are infrared divergences.
As in QED, these are regarded as `harmless' for calculation of
physical effects. I thank Ashoke Sen for
discussions on this issue.} %
However Gross and Periwal have shown that in the case of
bosonic strings, when summed, the series diverges and does
so uncontrollably. (Technically, it is not even
Borel-summable.) They also gave arguments that the
conclusion would not be changed if one uses superstrings
instead. Independent support for these arguments has come
from work on random surfaces due to Ambjorn and others. One
might wonder why the divergence of the sum should be
regarded as a serious failure of the theory. After all, in
quantum electrodynamics, the series is also believed to
diverge. Recall however that quantum electrodynamics is an
inherently incomplete theory. It ignores many processes
that come into play at high energies or short distances. In
particular, it completely ignores the microstructure of
space-time and simply assumes that space-time can be
approximated by a smooth continuum even below the Planck
scale. Therefore, it can plead incompleteness and shift the
burden of this infinity to a more complete theory. A
`theory of everything' on the other hand, has nowhere to
hide. It cannot plead incompleteness and shift its burden.
It must face the Planck regime squarely. If the theory is
to be consistent, it must have key non-perturbative
structures.

The current and the fourth stage of the particle physics
motivated approaches to quantum gravity is largely devoted to
unraveling such structures and using them to make unsuspected
connections between gravity and other areas of physics such as
fluid dynamics and high temperature superconductivity. It is
widely believed that the AdS/CFT conjecture provides a
non-perturbative definition of string theory on space-times
satisfying certain boundary conditions \cite{ghjprev}. More
precisely, in this correspondence string theory on
asymptotically AdS bulk space-times is taken to be equivalent to
certain gauge theories on its boundary. However, from the
perspective of quantum gravity, this approach has some serious
limitations. First, a negative cosmological constant in the bulk
is essential to this correspondence while the observed
cosmological constant is positive. Considerable effort were made
initially to extend the ideas to a positive or zero cosmological
constant but they have not had notable success. Second, the bulk
space-time is 10 dimensional. One can compactify the unwanted
dimensions using n-spheres, but the compactified directions
cannot be microscopic; the correspondence requires that the
radius of these spheres should equal the cosmological radius! So
if one just looks around, one should see these large macroscopic
dimensions. Consequently, the non-perturbative string theory
defined through the conjecture has little to do with the
macroscopic world we live in. Finally, even if one overlooks
this issue and considers space-times with higher macroscopic
dimensions, the AdS/CFT duality is yet to shed light on the
conceptually central issues such as the space-time structure
inside the horizon of an evaporating black hole or the fate of
big-bang type cosmological singularities. The recent thrust
---and the strength--- of these developments is, rather, that
they enable one to use known techniques from gravity and
supergravity to solve some of the difficult mathematical
problems encountered in the strong coupling regimes of field
theories describing \emph{non-gravitational} systems.

On the relativity side, the third stage began with the following
observation: the geometrodynamics program laid out by Dirac,
Bergmann, Wheeler and others simplifies significantly if we
regard a spatial connection ---rather than the 3-metric--- as
the basic object. In fact we now know that, among others,
Einstein and Schr\"odinger had recast general relativity as a
theory of connections already in the fifties.(For a brief
account of this fascinating history, see \cite{aa1}.) However,
they used the `Levi-Civita connection' that features in the
parallel transport of vectors and found that the theory becomes
rather complicated. This episode had been forgotten and
connections were re-introduced afresh in the mid 1980s
\cite{aabook}.%
\footnote{This reformulation used (anti-)self-dual connections
which are complex. These have a direct interpretation in terms
space-time geometry and also render the constraint equations
polynomial in the basic variables. This simplicity was regarded
as crucial for passage to quantum theory. However, one is then
faced with the task of imposing appropriate quantum `reality
conditions' to ensure that the classical limit is real general
relativity. Barbero introduced real connection variables by
replacing the $\pm i$ in the expression of the (anti-)self-dual
connections with a real parameter $\beta$. However, now the
connection does not have a natural space-time interpretation and
the constraints are no longer polynomial in the basic variables.
But the strategy became viable after Thiemann introduced novel
ideas to handle quantization of the specific non-polynomial
terms that now feature in the constraints. Since then this
strategy has become crucial because the rigorous functional
calculus on the space of connections has so far been developed
only for real connections. Immirzi suggested that the value of
$\beta$ could be chosen so that the leading term in black hole
entropy is precisely (area/$4\lp^2$). That is why $\beta$ (which
is often denoted by $\gamma$ in later papers) is referred to as
the \emph{Barbero-Immirzi} parameter.} 
%
However, now these are `spin-connections' required to parallel
propagate spinors, and they turn out to \emph{simplify}
Einstein's equations considerably. For example, the dynamical
evolution dictated by Einstein's equations can now be visualized
simply as a \emph{geodesic motion} on the space of
spin-connections (with respect to a natural metric extracted
from the constraint equations). Since general relativity is now
regarded as a dynamical theory of connections, this
reincarnation of the canonical approach is called
`connection-dynamics'.

Perhaps the most important advantage of the passage from metrics
to connections is that the phase-space of general relativity is
now the same as that of gauge theories \cite{aabook,jbbook}. The
`wedge between general relativity and the theory of elementary
particles' that Weinberg referred to largely disappears without
having to sacrifice the geometrical essence of general
relativity. One could now import into general relativity
techniques that have been highly successful in the quantization
of gauge theories. At the kinematic level, then, there is a
unified framework to describe all four fundamental interactions.
The dynamics, of course, depends on the interaction. In
particular, while there is a background space-time geometry in
electroweak and strong interactions, there is none in general
relativity. Therefore, qualitatively new features arise. These
were exploited in the late eighties and early nineties to solve
simpler models ---general relativity in 2+1 dimensions
\cite{aabook,aalh,sc}; linearized gravity clothed as a gauge
theory \cite{aabook}; and certain cosmological models. To
explore the physical, 3+1 dimensional theory, a `loop
representation' was introduced by Rovelli and Smolin. Here,
quantum states were taken to be
suitable functions of loops on the 3-manifold.%
\footnote{This is the origin of the name `loop quantum
gravity'. The loop representation played an important role
in the initial stages. Although this is no longer the case
in the current, fourth phase, the name is still used to
distinguish this approach from others.}
This led to a number of interesting and intriguing results,
particularly by Gambini, Pullin and their collaborators,
relating knot theory and quantum gravity \cite{gpbook}. Thus,
there was rapid and unanticipated progress in a number of
directions which rejuvenated the canonical quantization program.
Since the canonical approach does not require the introduction
of a background geometry or use of perturbation theory, and
because one now has access to fresh, non-perturbative techniques
from gauge theories, in relativity circles there is a hope that
this approach may lead to well-defined, \emph{non-perturbative}
quantum general relativity, or its supersymmetric version,
supergravity.

However, a number of these considerations remained rather
formal until mid-nineties. Passage to the loop
representation required an integration over the infinite
dimensional space of connections and the formal methods
were insensitive to possible infinities lurking in the
procedure. Indeed, such integrals are notoriously difficult
to perform in interacting field theories. To pay due
respect to the general covariance of Einstein's theory, one
needed diffeomorphism invariant measures and there were
folk-theorems to the effect that such measures did not
exist!

Fortunately, the folk-theorems turned out to be incorrect. To
construct a well-defined theory capable of handling field
theoretic issues, a \emph{quantum theory of Riemannian geometry}
was systematically constructed in the mid-nineties \cite{alrev}.
This launched the fourth (and the current) stage in the
canonical approach. Just as differential geometry provides the
basic mathematical framework to formulate modern gravitational
theories in the classical domain, quantum geometry provides the
necessary concepts and techniques in the quantum domain. It is a
rigorous mathematical theory which enables one to perform
integration on the space of connections for constructing Hilbert
spaces of states and to define geometric operators
corresponding, e.g. to areas of surfaces and volumes of regions,
even though the classical expressions of these quantities
involve non-polynomial functions of the Riemannian metric. There
are no infinities. One finds that, at the Planck scale, geometry
has a definite discrete structure. Its fundamental excitations
are 1-dimensional, rather like polymers, and the space-time
continuum arises only as a coarse-grained approximation. The
fact that the structure of space-time at Planck scale is
qualitatively different from Minkowski background used in
perturbative treatments reinforced the idea that quantum general
relativity (or supergravity) may well be non-perturbatively
finite. Quantum geometry effects have already been shown to
resolve the big-bang singularity and solve some of the
long-standing problems associated with black holes. (See
lectures by Giesel, Sahlmann and Singh at this School.)

Over the last six years, another frontier has advanced in loop
quantum gravity: \emph{spin foams} (and the associated
development of group field theory) which provide a sum over
histories formulation \cite{aprev,crbook,gft1}. The new element
here is that the histories that enter the sum are \emph{quantum}
geometries of a specific type; they can be regarded as the `time
evolution' of the polymer-like quantum 3 geometries that emerged
in the canonical approach. So far the sum has not been
systematically derived starting from the classical theory as one
generally does in, say, gauge theories. Rather, one uses
semi-heuristic considerations to arrive at a definition of the
`transition amplitudes' and then explores physical properties of
the resulting quantum theory. There are detailed arguments to
the effect that one recovers the Einstein Hilbert action in an
appropriate limit. Furthermore, although the underlying theory
is diffeomorphism invariant, given a suitable `boundary state',
there is a conceptual framework to calculate n-point functions
normally used in perturbative treatments. Information about the
background space-time on which these n-point functions live is
encoded in the chosen `boundary state'. However, a number of
important problems still remain. The status is described in
detail in the lectures by Rovelli, Speziale, Baratin, Perini,
Fairbairn, Bianchi and Kaminski at
this school.\\
\bigskip

The first three stages of developments in quantum gravity taught
us many valuable lessons. Perhaps the most important among them
is the realization that perturbative, field theoretic methods
which have been so successful in other branches of physics are
inadequate to construct quantum gravity. The assumption that
space-time can be replaced by a smooth continuum at arbitrarily
small scales leads to inconsistencies. We can neither ignore the
microstructure of space-time nor presuppose its nature. We must
let quantum gravity itself reveal this structure to us.

For brevity and to preserve the flow of discussion, I have
restricted myself to the `main-stream' programs whose
development can be continuously tracked over several decades.
However, I would like to emphasize that there are a number of
other highly original approaches ---particularly, the Euclidean
path integral approach \cite{wiswh}, Regge calculus
\cite{reggerev}, asymptotic safety scenarios \cite{reuterrev},
discrete approaches \cite{gp}, causal dynamical triangulations
\cite{lollrev,loll}, twistor theory \cite{rp1,rpwr,rp2} and the
theory of H-spaces \cite{hspace}, asymptotic quantization
\cite{aa3}, non-commutative geometry \cite{connes}, causal sets
\cite{sorkin,dowker} and Topos theory \cite{cji4,dahlen}. Ideas
underlying several of these approaches are inter-related and
some of them lie at the foundation of other avenues. This is
particularly true of the path integral approach pioneered by
Misner, and developed in much greater detail in the Euclidean
context by Hawking, Hartle, Halliwell and others. Ideas
developed in this approach have provided the point of departure
for the ongoing developments in causal dynamical triangulations,
asymptotic safety and spin foams.

\subsection{Physical Questions of Quantum Gravity}
 \label{s1.2}

Approaches to quantum gravity face two types of issues: Problems
that are `internal' to individual programs and physical and
conceptual questions that underlie the whole subject. Examples
of the former are: Incorporation of physical ---rather than half
flat--- gravitational fields in the twistor program, mechanisms
for breaking of supersymmetry and dimensional reduction in
string theory, and issues of space-time covariance in the
canonical approach. In this sub-section, I will focus on the
second type of issues by recalling some of the long standing
issues that \emph{any} satisfactory quantum theory of gravity
should address.\\

$\bullet$ \textit{The Big-Bang and other singularities}: It is
widely believed that the prediction of a singularity, such as
the big-bang of classical general relativity, is primarily a
signal that the physical theory has been pushed beyond the
domain of its validity. A key question to any quantum gravity
theory, then, is: What replaces the big-bang? Are the classical
geometry and the continuum picture only approximations,
analogous to the `mean (magnetization) field' of ferro-magnets?
If so, what are the microscopic constituents? What is the
space-time analog of a Heisenberg quantum model of a
ferro-magnet? When formulated in terms of these fundamental
constituents, is the evolution of the \textit{quantum} state of
the universe free of singularities? General relativity predicts
that the space-time curvature must grow unboundedly as we
approach the big-bang or the big-crunch but we expect the
quantum effects, ignored by general relativity, to intervene,
making quantum gravity indispensable before infinite curvatures
are reached. If so, what is the upper bound on curvature? How
close to the singularity can we `trust' classical general
relativity? What can we say about the `initial conditions',
i.e., the quantum state of geometry and matter that correctly
describes the big-bang? If they have to be imposed externally,
is there a \textit{physical} guiding principle?

 \b\textit{Black holes:} In the early seventies, using
imaginative thought experiments, Bekenstein argued that black
holes must carry an entropy proportional to their area
\cite{wiswh,waldrev,akrev}. About the same time, Bardeen, Carter
and Hawking (BCH) showed that black holes in equilibrium obey
two basic laws, which have the same form as the zeroth and the
first laws of thermodynamics, provided one equates the black
hole surface gravity $\kappa$ to some multiple of the
temperature $T$ in thermodynamics and the horizon area $a_{\rm
hor}$ to a corresponding multiple of the entropy $S$
\cite{wiswh,waldrev,akrev}. However, at first this similarity
was thought to be only a formal analogy because the BCH analysis
was based on \textit{classical} general relativity and simple
dimensional considerations show that the proportionality factors
must involve Planck's constant $\hbar$. Two years later, using
quantum field theory on a black hole background space-time,
Hawking showed that black holes in fact radiate quantum
mechanically as though they are black bodies at temperature $T =
\hbar\kappa/2\pi$ \cite{wiswh,waldbook}. Using the analogy with
the first law, one can then conclude that the black hole entropy
should be given by $S_{\rm BH} = a_{\rm hor}/4G\hbar$. This
conclusion is striking and deep because it brings together the
three pillars of fundamental physics
---general relativity, quantum theory and statistical mechanics.
However, the argument itself is a rather hodge-podge mixture of
classical and semi-classical ideas, reminiscent of the Bohr
theory of atom. A natural question then is: what is the analog
of the more fundamental, Pauli-Schr\"odinger theory of the
Hydrogen atom? More precisely, what is the statistical
mechanical origin of black hole entropy? What is the nature of a
quantum black hole and what is the interplay between the quantum
degrees of freedom responsible for entropy and the exterior
curved geometry? Can one derive the Hawking effect from first
principles of quantum gravity? Is there an imprint of the
classical singularity on the final quantum description, e.g.,
through `information loss'?

\b \textit{Planck scale physics and the low energy world:} In
general relativity, there is no background metric, no inert
stage on which dynamics unfolds. Geometry itself is dynamical.
Therefore, as indicated above, one expects that a fully
satisfactory quantum gravity theory would also be free of a
background space-time geometry. However, of necessity, a
background independent description must use physical concepts
and mathematical tools that are quite different from those of
the familiar, low energy physics. A major challenge then is to
show that this low energy description does arise from the
pristine, Planckian world in an appropriate sense, bridging the
vast gap of some 16 orders of magnitude in the energy scale. In
this `top-down' approach, does the fundamental theory admit a
`sufficient number' of semi-classical states? Do these
semi-classical sectors provide enough of a background geometry
to anchor low energy physics? Can one recover the familiar
description? If the answers to these questions are in the
affirmative, can one pin point why the standard `bottom-up'
perturbative approach fails? That is, what is the essential
feature which makes the fundamental description mathematically
coherent but is absent in the standard perturbative quantum
gravity?\\

There are of course many more challenges: adequacy of
standard quantum mechanics, the issue of time, of
measurement theory and the associated questions of
interpretation of the quantum framework, the issue of
diffeomorphism invariant observables and practical methods
of computing their properties, convenient ways of computing
time evolution and S-matrices, exploration of the role of
topology and topology change, etc, etc. In loop quantum
gravity described in the rest of this chapter, one adopts
the view that the three issues discussed in detail are more
basic from a physical viewpoint because they are rooted in
general conceptual questions that are largely independent
of the specific approach being pursued. Indeed they have
been with us longer than any of the current leading
approaches.

\section{Loop quantum gravity}
\label{s2}

In this section, I will summarize the overall viewpoint,
achievements, challenges and opportunities underlying loop
quantum gravity. The emphasis is on structural and conceptual
issues. Detailed treatments of the subject can be found in
lectures by Giesel, Sahlmann, Rovelli and Singh in these
proceedings and even more complete and more technical accounts
in \cite{alrev,crbook,ttbook} and references therein. (The
development of the subject can be seen by following older
monographs \cite{aabook,jbbook,gpbook}.) For a treatment at a
more elementary (i.e. advanced undergraduate) level, see
\cite{gp-book}.

\subsection{Viewpoint}
\label{s2.1}

In loop quantum gravity, one takes the central lesson of
general relativity seriously: gravity \textit{is} geometry
whence, in a fundamental quantum gravity theory, there
should be no background metric. Geometry and matter should
\textit{both} be `born quantum mechanically'. Thus, in
contrast to approaches developed by particle physicists,
one does not begin with quantum matter on a background
geometry and use perturbation theory to incorporate quantum
effects of gravity. There \textit{is} a manifold but no
metric, or indeed any other physical fields, in the background.%
\footnote{In 2+1 dimensions, although one begins in a
completely analogous fashion, in the final picture one can
get rid of the background manifold as well. Thus, the
fundamental theory can be formulated combinatorially
\cite{aabook,aalh}. While some steps have been taken to
achieve this in 3+1 dimensions, by considering `abstract'
spin networks in the canonical approach and 2-complexes in
spin foams, one still needs a more complete handle on the
underlying mathematics.}

In classical gravity, Riemannian geometry provides the
appropriate mathematical language to formulate the physical,
kinematical notions as well as the final dynamical equations.
This role is now taken by \textit{quantum} Riemannian geometry.
In the classical domain, general relativity stands out as the
best available theory of gravity, some of whose predictions have
been tested to an amazing degree of accuracy, surpassing even
the legendary tests of quantum electrodynamics. Therefore, it is
natural to ask: \textit{Does quantum general relativity, coupled
to suitable matter} (or supergravity, its supersymmetric
generalization) \textit{exist as consistent theories
non-perturbatively?} There is no implication that such a theory
would be the final, complete description of Nature. Nonetheless,
this is a fascinating and important open question in its own
right.

As explained in section \ref{s1.1}, in particle physics
circles the answer is often assumed to be in the negative,
not because there is concrete evidence against
non-perturbative quantum gravity, but because of the
analogy to the theory of weak interactions. There, one
first had a 4-point interaction model due to Fermi which
works quite well at low energies but which fails to be
renormalizable. Progress occurred not by looking for
non-perturbative formulations of the Fermi model but by
replacing the model by the Glashow-Salam-Weinberg
renormalizable theory of electro-weak interactions, in
which the 4-point interaction is replaced by $W^\pm$ and
$Z$ propagators. Therefore, it is often assumed that
perturbative non-renormalizability of quantum general
relativity points in a similar direction. However this
argument overlooks the crucial fact that, in the case of
general relativity, there is a qualitatively new element.
Perturbative treatments pre-suppose that the space-time can
be assumed to be a continuum \textit{at all scales} of
interest to physics under consideration. This assumption is
safe for weak interactions. In the gravitational case, on
the other hand, the scale of interest is \emph{the Planck
length} $\lp$ and there is no physical basis to pre-suppose
that the continuum picture should be valid down to that
scale. The failure of the standard perturbative treatments
may largely be due to this grossly incorrect assumption and
a non-perturbative treatment which correctly incorporates
the physical micro-structure of geometry may well be free
of these inconsistencies.

Are there any situations, outside loop quantum gravity, where
such physical expectations are borne out in detail
mathematically? The answer is in the affirmative. There exist
quantum field theories (such as the Gross-Neveu model in three
dimensions) in which the standard perturbation expansion is not
renormalizable although the theory is \emph{exactly soluble}!
Failure of the standard perturbation expansion can occur because
one insists on perturbing around the trivial, Gaussian point
rather than the more physical, non-trivial fixed point of the
renormalization group flow. Interestingly, thanks to recent work
by Reuter, Lauscher, Percacci, Perini and others there is now
non-trivial and growing evidence that situation may be similar
in Euclidean quantum gravity. Impressive calculations have shown
that pure Einstein theory may also admit a non-trivial fixed
point \cite{reuterrev,reuter}. Furthermore, the requirement that
the fixed point should continue to exist in presence of matter
constrains the couplings in non-trivial and interesting ways
\cite{perini}.

However, as indicated in the Introduction, even if quantum
general relativity did exist as a mathematically consistent
theory, there is no a priori reason to assume that it would
be the `final' theory of all known physics. In particular,
as is the case with classical general relativity, while
requirements of background independence and general
covariance do restrict the form of interactions between
gravity and matter fields and among matter fields
themselves, the theory would not have a built-in principle
which \textit{determines} these interactions. Put
differently, such a theory would not be a satisfactory
candidate for unification of all known forces. However,
just as general relativity has had powerful implications in
spite of this limitation in the classical domain, quantum
general relativity should have qualitatively new
predictions, pushing further the existing frontiers of
physics. Indeed, unification does not appear to be an
essential criterion for usefulness of a theory even in
other interactions. QCD, for example, is a powerful theory
even though it does not unify strong interactions with
electro-weak ones. Furthermore, the fact that we do not yet
have a viable candidate for the grand unified theory does
not make QCD any less useful.

\subsection{Advances}
 \label{s2.2}

{}From the historical and conceptual perspectives of section
\ref{s1}, loop quantum gravity has had several successes. Thanks
to the systematic development of quantum geometry, several of
the roadblocks encountered by quantum geometrodynamics were
removed. Functional analytic issues related to the presence of
an infinite number of degrees of freedom are now faced squarely.
Integrals on infinite dimensional spaces are rigorously defined
and the required operators have been systematically constructed.
Thanks to this high level of mathematical precision, the
Hamiltonian and the spin foam programs in loop quantum gravity
have leaped past the `formal' stage of development. More
importantly, although key issues related to quantum dynamics
still remain, it has been possible to use the parts of the
program that are already well established to extract useful and
highly non-trivial physical predictions. In particular, some of
the long standing issues about the nature of the big-bang,
physics of the very early universe, properties of quantum black
holes, giving meaning to the n-point functions in a background
independent framework have been resolved. In this sub-section, I
will further clarify some conceptual issues and discuss some
recent advances.\\

\b \emph{Quantum geometry.} The specific quantum Riemannian
geometry underlying loop quantum gravity predicts that
eigenvalues of geometric operators ---such as areas of
2-surfaces and volumes of 3-dimensional regions--- are discrete.
Thus, continuum underlying general relativity is only a coarse
grained approximation. What is the direct \emph{physical}
significance of this specific discreteness? Recall first that,
in the classical theory, differential geometry simply provides
us with formulas to compute areas of surfaces and volumes of
regions in a Riemannian manifold. To turn these quantities into
physical observables of general relativity, one has to define
the surfaces and regions \emph{operationally}, e.g. by focusing
on surfaces of black holes or regions in which matter fields are
non-zero. Once this is done, one can simply use the formulas
supplied by differential geometry to calculate values of these
observable. The situation is rather similar in loop quantum
gravity. For instance, the area of the isolated horizon is a
Dirac observable in the classical theory and the application of
the quantum geometry area formula to \emph{this} surface leads
to physical results. In 2+1 dimensions, Freidel, Noui and Perez
have introduced point particles coupled to gravity. The physical
distance between these particles is again a Dirac observable.
When used in this context, the spectrum of the length operator
has direct physical meaning. In all these situations, the
operators and their eigenvalues correspond to the `proper'
lengths, areas and volumes of physical objects, measured in the
rest frames. Finally sometimes questions are raised about
compatibility between discreteness of these eigenvalues and
Lorentz invariance. As was emphasized by Rovelli and Speziale,
there is no tension whatsoever: it suffices to recall that
discreteness of eigenvalues of the angular momentum operator
$\hat{J}_z$ of non-relativistic quantum mechanics is perfectly
compatible with the rotational invariance of that theory.\\

\b  \emph{Quantum cosmology.} In
Friedmann-Lemaitre-Robertson-Walker (FLRW) models, loop quantum
gravity has resolved the long-standing physical problem of the
fate of the big-bang in quantum gravity \cite{asrev}. Work by
Bojowald, Ashtekar, Pawlowski, Singh and others has shown that
non-perturbative effects originating in quantum geometry create
an effective repulsive force which is negligible when the
curvature falls significantly below the Planck scale but rises
very quickly and dramatically in the deep Planck regime to
overcome the classical gravitational attraction, thereby
replacing the big-bang by a quantum bounce. The same is true
with the big-crunch singularity in the closed models. More
generally, using effective equations, Singh has shown that these
quantum geometry effects also resolve \emph{all} strong
curvature singularities in homogeneous isotropic models where
matter sources have an equation of state of the type
$p=p(\rho)$, including the exotic singularities such as the
big-rip. (These can occur with non-standard matter, still
described by an equation of state $p = p(\rho)$).

A proper treatment of anisotropies (i.e. Bianchi models) has
long been a highly non-trivial issue in general bouncing
scenarios because the anisotropic shears dominate in Einstein's
equations in the contracting phase before the bounce, diverging
(as $1/a^6$ which is) faster than, say, the dust or radiation
matter density. Therefore, if anisotropies are added even as a
perturbation to a FLRW model, they tend to grow unboundedly.
What is the situation in loop quantum cosmology? The issue
turned out to be quite subtle and there were some oversights at
first. But a careful examination by Ashtekar, Wilson-Ewing and
others has shown that the singularity is again resolved: any
time a shear scalar ---a potential for the Weyl curvature--- or
matter density approaches the Planck regime, the repulsive force
of quantum geometry grows to dilute it. As in the isotropic
case, effective equations can again be used to gain physical
insights. In particular they show that the matter density is
again bounded above. Singularity resolution in these Bianchi
models is also important from a more general consideration.
There is a conjecture due to Belinskii, Khalatnikov and Lifshitz
(BKL) that says that as one approaches a space-like singularity
in classical general relativity, `the terms containing time
derivatives dominate over those containing spatial derivatives,
so that the dynamics of the gravitational field at any one
spatial point are better and better approximated by the dynamics
of Bianchi models'. By now considerable evidence has accumulated
in support of the BKL conjecture and it is widely believed to be
essentially correct. One might therefore hope that the
singularity resolution in the Bianchi models in loop quantum
cosmology has opened a door to showing that all strong
curvature, space-like singularities are resolved by the quantum
geometry effects underlying loop quantum gravity.

Finally, the simplest type of (non-linear) inhomogeneous models
---the 1-polarization Gowdy space-times--- have also been analyzed
in detail. These models were studied extensively in the early
quantum gravity literature, prior to the advent of LQC. In all
cases the singularity had persisted. A systematic study in the
context of loop quantum cosmology was initiated by Mena,
Martin-Benito, Pawlowski and others by making an astute use of
the fact that the homogeneous modes of the model correspond to a
Bianchi I space-time. Once again, the underlying quantum
geometry resolves the big-bang singularity.

I will conclude with the discussion of a conceptual point. In
general relativity, non-singular, bouncing models can be and
have been constructed by using matter fields that violate energy
conditions. In loop quantum cosmology, by contrast, matter
fields satisfy all energy conditions. How can the theory then
evade singularity theorems of Penrose, Hawking and others? It
does so because the quantum geometry effects modify the
geometric, left hand side of Einstein's equations, whence these
theorems are inapplicable. However there are more recent
singularity theorems due to Borde, Guth and Vilenkin which do
\emph{not} refer to field equations at all. How are these
evaded? These theorems were motivated by inflationary scenario
and therefore assume that the universe bas been eternally
undergoing an expansion. In loop quantum cosmology, even with an
inflationary potential, the pre-bounce branch is contracting.
Thus again the singularity is avoided because the solutions
violate a key assumption of these theorems as well.\\

\b  \emph{Quantum Horizons.} Loop quantum cosmology
illuminates dynamical ramifications of quantum geometry but
within the context of mini and midi superspaces where an
infinite number of degrees of freedom are frozen. The
application to the black hole entropy problem is
complementary in that one considers the full theory but
probes consequences of quantum geometry which are not
sensitive to full quantum dynamics. I will discuss this
topic in a little more detail because it was not covered in
any of the main lectures at this school.

As explained in the Introduction, since mid-seventies, a key
question in the subject has been: What is the statistical
mechanical origin of the entropy  $S_{\rm BH} = ({a_{\rm hor}/
4\lp^2})$ of large black holes? What are the microscopic degrees
of freedom that account for this entropy? This relation implies
that a solar mass black hole must have\, $\exp 10^{77}$\,
quantum states, a number that is \textit{huge} even by the
standards of statistical mechanics. Where do all these states
reside? To answer these questions, in the early 1990s Wheeler
had suggested the following heuristic picture, which he
christened `It from Bit'. Divide the black hole horizon into
elementary cells, each with one Planck unit of area, $\lp^2$ and
assign to each cell two microstates, or one `Bit'. Then the
total number of states ${\cal N}$ is given by ${\cal N} = 2^n$
where $n = ({a_{\rm hor}/ \lp^2})$ is the number of elementary
cells, whence entropy is given by $S = \ln {\cal N} \sim a_{\rm
hor}$. Thus, apart from a numerical coefficient, the entropy
(`It') is accounted for by assigning two states (`Bit') to each
elementary cell. This qualitative picture is simple and
attractive. But can these heuristic ideas be supported by a
systematic analysis from first principles?

Ashtekar, Baez, Corichi and Krasnov used quantum geometry to
provide such an analysis. The first step was to analyze the
structure of `isolated horizons' in general relativity
\cite{akrev} and use it in conjunction to quantum geometry to
define an isolated \emph{quantum horizon}. To probe its
properties, one has to combine the isolated horizon boundary
conditions from classical general relativity and quantum
Riemannian geometry of loop quantum gravity with the
Chern-Simons theory on a punctured sphere, the theory of a
non-commutative torus and subtle considerations involving
mapping class groups. This detailed analysis showed that, while
qualitative features of Wheeler's picture are borne out,
geometry of a quantum horizon is much more subtle. First, while
Wheeler's ideas hold for any 2-surface, the loop quantum gravity
calculation requires a quantum horizon. Second, basic features
of both of Wheeler's arguments undergo a change: i) the
elementary cells do not have Planck area; values of their area
are dictated by the spectrum, $\sim \sqrt{j(j+1)}$, of the area
operator in loop quantum gravity, where $j$ is a half integer;
ii) individual cells carry much more than just one `bit' of
information; the number of states associated with any one cell
is $2j+1$.

Nonetheless, a careful counting of states by Lewandowski,
Domagala, Meissner and others has shown that the number of
microstates is again proportional to the area of the isolated
horizon. To get the exact numerical factor of $1/4$, one has to
fix the Barbero-Immirzi parameter of loop quantum gravity to a
specific value. One can use a specific type of isolated horizon
for this ---e.g. the spherically symmetric one with zero charge,
or the cosmological one in the de Sitter space-time. Once the
value of the parameter is fixed, one gets the correct numerical
coefficient in the leading order contribution for isolated
horizons with arbitrary mass and angular momentum moments,
charge, etc. (One also obtains a precise logarithmic sub-leading
correction, whose coefficient does not depend on the
Barbero-Immirzi parameter.) The final result has two significant
differences with respect to the string theory calculations: i)
one does not require near-extremality; one can handle ordinary
4-dimensional black holes of direct astrophysical interest which
may be distorted and/or rotating; and, ii) one can
simultaneously incorporate cosmological horizons for which
thermodynamics considerations also apply \cite{wiswh}.

Why does this value of the Barbero-Immirzi parameter not
depend on non-gravitational charges? This important
property can be traced back to a key consequence of the
isolated horizon boundary conditions: detailed calculations
show that only the gravitational part of the symplectic
structure has a surface term at the horizon; the matter
symplectic structures have only volume terms. (Furthermore,
the gravitational surface term is insensitive to the value
of the cosmological constant.) Consequently, there are no
independent surface quantum states associated with matter.
This provides a natural explanation of the fact that the
Hawking-Bekenstein entropy depends only on the horizon area
and is independent of electro-magnetic (or other) charges.
(For more detailed accounts of these results, see
\cite{alrev,akrev}.)

Over the last three years there has been a resurgence of
interest in the subject, thanks to the impressive use of number
theory techniques by Barbero, Villasenor, Agullo, Borja,
Diaz-Polo and to sharpen and very significantly extend the
counting of horizon states. These techniques have opened new
avenues to further explore the microstates of the quantum
horizon geometry through contributions by Perez, Engle, Noui,
Pranzetti, Ghosh, Mitra, Kaul, Majumdar and others.

To summarize, as in other approaches to black hole entropy,
concrete progress could be made in loop quantum gravity because:
i) the analysis does not require detailed knowledge of how
quantum dynamics is implemented in \emph{full} theory, and, ii)
restriction to large black holes implies that the Hawking
radiation is negligible, whence the black hole surface can be
modeled by an isolated horizon \cite{akrev}. The states
responsible for entropy have a direct interpretation in
\emph{space-time} terms: they refer to the geometry of the
quantum, isolated horizon.\\

\b  \emph{Quantum Einstein's equations in the canonical
framework.} The challenge of quantum dynamics in the full theory
is to find solutions to the quantum constraint equations and
endow these physical states with the structure of an appropriate
Hilbert space. The general consensus in the loop quantum gravity
community is that while the situation is well-understood for the
Gauss and diffeomorphism constraints, it is far from being
definitive for the Hamiltonian constraint. Non-trivial
development due to Thiemann is that well-defined candidate
operators representing the Hamiltonian constraint exist on the
space of solutions to the Gauss and diffeomorphism constraints
\cite{ttbook}. However there are several ambiguities
\cite{alrev} and, unfortunately, we do not understand the
physical meaning of choices made to resolve them. Detailed
analysis in the limited context of loop quantum cosmology has
shown that choices which appear to be mathematically natural can
nonetheless lead to unacceptable physical consequences such as
departures from general relativity in completely tame situations
with low curvature \cite{asrev}. Therefore, much more work is
needed in the full theory.

The current status can be summarized as follows. Four main
avenues have been pursued to construct and solve the quantum
Hamiltonian constraint. The first is the `Master constraint
program' introduced by Thiemann \cite{ttbook}. The idea here is
to avoid using an infinite number of Hamiltonian constraints
${\cal S}(N) = \int N(x) {\cal{S}}(x) d^3x$, each smeared by a
so-called `lapse function' $N$. Instead, one squares the
integrand ${\cal S}(x)$ itself in an appropriate sense and then
integrates it on the 3-manifold $M$. In simple examples, this
procedure leads to physically viable quantum theories. However,
in loop quantum gravity the procedure does not remove any of the
ambiguities in the definition of the Hamiltonian constraint.
Rather, if the ambiguities are resolved, the principal strength
of the strategy lies in its potential to complete the last step
in quantum dynamics: finding the physically appropriate scalar
product on physical states. The general philosophy is similar to
that advocated by John Klauder over the years in his approach to
quantum gravity based on coherent states \cite{klauder}. A
second strategy to solve the quantum Hamiltonian constraint is
due to Gambini, Pullin and their collaborators. It builds on
their extensive work on the interplay between quantum gravity
and knot theory \cite{gpbook}. The more recent of these
developments use the relatively new invariants of
\emph{intersecting} knots discovered by Vassiliev. This is a
novel approach which furthermore has a potential of enhancing
the relation between topological field theories and quantum
gravity. As our knowledge of invariants of intersecting knots
deepens, this approach could provide increasingly significant
insights. In particular, it has the potential of leading to a
formulation of quantum gravity which does not refer even to a
background manifold (see footnote 9).

The third approach comes from spin-foam models
\cite{aprev,crbook} which, as discussed below, provide a path
integral approach to quantum gravity. Over the last four years,
there has been extensive work in this area, discussed in the
articles by Rovelli, Speziale, Baratin, Perini, Fairbairn,
Bianchi, and Kaminski in this volume. Transition amplitudes from
path integrals can be used to restrict the choice of the
Hamiltonian constraint operator in the canonical theory. This is
a very promising direction and Freidel, Noui, Perez, Rovelli and
others have analyzed this issue especially in 2+1 dimensions.
The idea in the fourth approach, due to Varadarajan, Laddha,
Henderson, Tomlin and others, is to use insights gained from the
analysis of parameterized field theories. Now the emphasis is on
drastically reducing the large freedom in the choice of the
definition of the Hamiltonian constraint by requiring that the
quantum constraint algebra closes, so that one is assured that
there is no obstruction to obtaining a large number of
\emph{simultaneous} solutions to all constraints. Because the
Poisson bracket between two Hamiltonian constraints is a
diffeomorphism constraint, one has to find a viable expression
of the operator generating \emph{infinitesimal} diffeomorphisms.
(Until this work, the focus was on the action only of
\emph{finite} diffeomorphisms in the kinematical setup.) Very
recently, this program has witnessed promising advances. The
Hamiltonian constraint one is led to define shares qualitative
features of `improved dynamics' of loop quantum cosmology that
lies at the foundation of the most significant advances in that
area.

In this discussion I have focused primarily on pure gravity. In
the mid 1990s Brown, Kucha$\check{\rm r}$ and Romano had
introduced frameworks in which matter fields can be used as
`rods and clocks' thereby providing a natural
`de-parametrization' of the constraints in the classical theory.
Giesel, Thiemann, Tamburino, Domagala, Kaminski, Lewandowski,
Husain and Pawlowski  have used these considerations as the
point of departure to construct loop quantum gravity theories
for these systems. Deparametrization greatly facilitates the
task of finding Dirac observables and makes it easier to
interpret the quantum theory. However, as in the Master
Constraint program, issues associated with quantization
ambiguities still remain and there is the domain on which matter
fields serve as good clocks and rods still needs to be
clarified. Further details can be found in the lectures by
Giesel and Salhmann. \\

\b \emph{Spin foams:} Four different avenues to quantum gravity
have been used to arrive at spin-foam models (SFMs). The fact
that ideas from seemingly unrelated directions converge to the
same type of structures and models has provided a strong impetus
to the spin foam program. Indeed, currently this is the most
active area on the mathematical physics side of loop quantum
gravity.

The first avenue is the Hamiltonian approach to loop quantum
gravity \cite{alrev,crbook,ttbook}. By mimicking the procedure
that led Feynman \cite{rpf} to a sum over histories formulation
of quantum mechanics, Rovelli and Reisenberger proposed a
space-time formulation of this theory. This work launched the
spin-foam program. The second route stems from the fact that the
starting point in canonical loop quantum gravity is a rewriting
of classical general relativity that emphasizes connections over
metrics \cite{alrev}. Therefore in the passage to quantum theory
it is natural to begin with the path integral formulation of
appropriate gauge theories. A particularly natural candidate is
the topological B-F theory because in 3 space-time dimensions it
is equivalent to Einstein gravity, and in higher dimensions
general relativity can be regarded as a constrained BF theory
\cite{jb,aprev}. The well-controlled path integral formulation
of the BF theory provided the second avenue and led to the SFM
of Barrett and Crane.

The third route comes from the Ponzano-Regge model of
3-dimensional gravity that inspired Regge calculus in higher
dimensions. Here one begins with a simplicial decomposition of
the space-time manifold, describes its discrete Riemannian
geometry using edge lengths and deficit angles and constructs a
path integral in terms of them. If one uses holonomies and
discrete areas of loop quantum gravity in place of edge lengths,
one is again led to a spin foam. These three routes are inspired
by various aspects of general relativity. The fourth avenue
starts from approaches to quantum gravity in which gravity is to
emerge from a more fundamental theory based on abstract
structures that, to begin with, have nothing to do with
space-time geometry. Examples are matrix models for
2-dimensional gravity and their extension to 3-dimensions
---the Boulatov model --- where the basic object is a field
on a group manifold rather than a matrix. The Boulatov model was
further generalized to a group field theory tailored to
4-dimensional gravity \cite{crbook,gft1}. The perturbative
expansion of this group field theory turned out be very closely
related to `vertex expansions' in SFMs. Thus the SFMs lie at a
junction where four apparently distinct paths to quantum gravity
meet. Through contributions of many researchers it has now
become an active research area (see, e.g., \cite{aprev,crbook}).

Four years ago, two groups, Engle-Livine-Pereira-Rovelli, and
Freidel-Krasnov, put forward precise proposals for the sum over
quantum geometries that could provide detailed dynamics in loop
quantum gravity. The motivations were different but for the
physically interesting values of the Babero-Immirzi parameter
(selected, e.g., by the black hole entropy considerations), the
two proposals agree. This is an improvement over the earlier
Barrett-Crane model which cured some of the problems faced by
that model. Perhaps more importantly, thanks to the
generalizations by Kaminski, Kisielowski and Lewandowski, the
canonical and path integral approaches have been brought closer
to one another: they use the same kinematics. However, there
does not yet exit a systematic `derivation' leading to this
proposal starting from classical general relativity, say, along
the lines used in textbooks to arrive at the path integral
formulation of gauge theories. Nonetheless the program has
attracted a large number of researchers because: i) there do
exist semi-heuristic considerations motivating the passage; ii)
as I indicated above, it can be arrived at from four different
avenues; and iii) Detailed asymptotic analysis by Barrett,
Hellmann, Dowdall, Fairbairn, Pereira and others strongly
indicates that these models have the correct classical limit;
and, iv) Because of the use of quantum geometry ---more
precisely because there is a non-zero area gap--- this sum over
quantum geometries has no ultra-violet divergences. More
recently, Fairbairn, Meusberger, Han and others have extended
these considerations to include a cosmological constant by a
natural use of quantum groups. It is then argued that, for a
given 2-simplex, the sum is also infra-red finite.

However, the issue of whether to sum over distinct 2-complexes
or to take an appropriate `continuum limit' is still debated and
it is not known whether the final result would be finite in
either case.%
\footnote{Mathematically, this situation is somewhat reminiscent
to perturbative super-string theory, where there is evidence
that each term in the expansion is finite but the sum is not
controlled.}
In the cosmological mini-superspaces, the situation is
well-controlled: under a single assumption that a sum and
an integral can be interchanged, the analog of the sum over
2-complexes (called the vertex expansion in the spin foam
literature) has been shown to converge, and furthermore,
converge to the `correct' result that is already known from
a well-established Hamiltonian theory \cite{asrev}.

A much more detailed discussion of spin foams can be found
in the lectures by Rovelli, Speziale, Baratin, Perini,
Fairbairn, Bianchi, and Kaminski at this school.

\subsection{Challenges and Opportunities}
\label{s2.3}

Developments summarized so far should suffice to provide a sense
of the extent to which advances in loop quantum gravity already
provide an avenue to a non-perturbative and background
independent formulation of quantum gravity. I will conclude by
providing an illustrative list of the open issues. Some of these
are currently driving the field while others provide challenges
and opportunities for further work. This discussion assumes that
the reader is familiar with basic ideas behind current research
in loop quantum gravity.

\subsubsection{Foundations}

\b \emph{Hamiltonian Theory:} In section \ref{s2.2} I outlined
four strategies that are being used to extract quantum dynamics.
I will now sketch another avenue, inspired in part by the
success of loop quantum cosmology, that has been proposed by
Domagala, Giesel, Kaminski and Lewandowski. In loop quantum
cosmology, a massless scalar field often serves as an `internal
clock' with respect to which observables of physical interest
evolve \cite{asrev}. The idea is to take over this strategy to
full quantum gravity by focusing on general relativity coupled
with a massless scalar field. This is a particularly interesting
system because, already in the 1990s, Kucha$\check{\rm r}$ and
Romano showed that one can rearrange the constraints of this
system so that they form a true Lie algebra, where the
\emph{Hamiltonian constraints Poisson commute with each other on
the entire phase space}. Interestingly, under seemingly mild
assumptions one can show that solutions of this system admit
space-like foliations on which $\phi$ is constant. Consequently,
even though the system has infinitely many degrees of freedom,
\emph{as in LQC} one can use $\phi$ as a relational time
variable. With $\mathbb{T}^3$ spatial topology for definiteness,
one can decompose all fields into homogeneous and \emph{purely}
inhomogeneous modes. If one were to truncate the system by
setting the inhomogeneous modes to zero, the resulting quantum
theory would be precisely the loop quantum cosmology of Bianchi
I models that has been analyzed in detail by Ashtekar and
Wilson-Ewing. One might imagine incorporating the inhomogeneous
modes using the `hybrid' quantization scheme that has been
successfully used in the Gowdy models by Mena, Martin-Benito,
Pawlowski and others, although it will have to be non-trivially
generalized to handle the fact that there are no Killing fields.
As in the Gowdy models, this will likely involve some gauge
fixing of the diffeomorphism and Gauss constraints. Even if
these gauge fixing strategies do not work globally on the full
phase space, one should still obtain a quantum theory tailored
to a `non-linear neighborhood' of FLRW or Bianchi I space-times.
Finally, effective equations for this system would also provide
valuable insights into the singularity resolution (which we
expect to persist in an appropriate, well-defined sense). In
particular, one would be able to compare and contrast their
prediction with the simple BKL behavior near the general
relativistic singularity, found by Andersson and Rendall for
this system. More generally, this analysis will enable one to
place loop quantum cosmology in the setting of full loop quantum
gravity.

A second important open issue is to find restrictions on matter
fields and their couplings to gravity for which this
non-perturbative quantization can be carried out to a
satisfactory conclusion. Supersymmetry, for example, is known to
allow only very specific matter content. Recent work by
Bodendorfer, Thiemann and Thurn has opened a fresh window for
this analysis. A second possibility is suggested by the analysis
of the closure of the constraint algebra in quantum theory. When
it is extended to allow for matter couplings, the recent work by
Varadarajan, Laddha, and Tomlin referred to in section\ref{s2.2}
could provide a promising approach to explore this issue in
detail. Finally, as mentioned in section \ref{s1.1}, the
renormalization group approach has provided interesting hints.
Specifically, Reuter et al have presented significant evidence
for a non-trivial fixed point for vacuum general relativity in 4
dimensions \cite{reuterrev}. When matter sources are included,
it continues to exist only when the matter content and couplings
are suitably restricted. For scalar fields, in particular,
Percacci and Perini have found that polynomial couplings (beyond
the quadratic term in the action) are ruled out, an intriguing
result that may `explain' the triviality of such theories in
Minkowski space-times \cite{perini}. Are there similar
constraints coming from loop quantum gravity?\\

\b \emph{Spin foams:} As discussed in section \ref{s2.2}, the
spin foam program has made significant advances over the last
four years. Results on the classical limit and finiteness of the
sum over histories for a fixed 2-complex are especially
encouraging. Therefore it is now appropriate to invest time and
effort on key foundational issues.

First, we need a better understanding of the physical meaning of
the `vertex expansion' that results when one sums over arbitrary
2-complexes. In particular, is there a systematic physical
approximation that lets us terminate the sum after a finite
number of terms? In group field theory each term is multiplied
by a power of the coupling constant \cite{crbook,gft1} but the
physical meaning of this coupling constant in space-time terms
is not known. Analysis by Ashtekar, Campiglia and Henderson in
the cosmological context bears out an early suggestion of Oriti
that the coupling constant is related to the cosmological
constant. In the full theory, Fairbairn, Meusberger, Han and
others have shown that the cosmological constant can be
incorporated using quantum groups (which also makes the spin
foam sum infrared finite for a fixed 2-complex). It is then
natural to ask if there is a precise sense in interpreting the
vertex expansion as a perturbation series in a parameter
physically related to the cosmological constant also in the full
theory.

Second, as I mentioned in section \ref{s2.2}, the issue of
whether one should actually sum over various 2-complexes (i.e.,
add up all terms in the vertex expansion), or take an
appropriately defined continuum limit is still open. Rovelli and
Smerlak have argued that there is a precise sense in which the
two procedures coincide. But so far there is no control over the
sum and experts in rigorous field theory have expressed the
concern that, unless a new principle is invoked, the number of
terms may grow uncontrollably as one increases the number of
vertices. Recent work on group field theory by Oriti, Rivasseau,
Gurau, Krajewski and others may help streamline this analysis
and provide the necessary mathematical control.

Finally, because the EPRL and FK models are motivated from
the BF theory, they inherit certain (`Plebanski') sectors
which classically do not correspond to general relativity.
In addition, analysis of cosmological spin foams
re-enforces an early idea due to Oriti that one should only
sum over `time oriented' quantum geometries. Some of these
issues are now being analyzed in detail by Engle and
others. But more work is needed on these basic conceptual
issues.\\

\b  \emph{Low energy physics:} In low energy physics one uses
quantum field theory on given background space-times. Therefore
one is naturally led to ask if this theory can be arrived at by
starting from loop quantum gravity and making systematic
approximations. Here, a number of interesting challenges appear
to be within reach. Fock states have been isolated in the
polymer framework \cite{alrev} and elements of quantum field
theory on quantum geometry have been introduced \cite{ttbook}.
These developments lead to concrete questions. For example, in
quantum field theory in flat space-times, the Hamiltonian and
other operators are regularized through normal ordering. For
quantum field theory on quantum geometry, on the other hand, the
Hamiltonians are expected to be manifestly finite
\cite{ttbook,alrev}. Can one then show that, in a suitable
approximation, normal ordered operators in the Minkowski
continuum arise naturally from these finite operators? Can one
`explain' why the so-called Hadamard states of quantum field
theory in curved space-times are special? These considerations
could also provide valuable hints for the construction of viable
semi-classical states of quantum geometry.

Since quantum field theory in FLRW space-times plays such an
important role in the physics of the early universe, it is
especially important to know if can be systematically derived
from loop quantum gravity. A number of obstacles immediately
come to mind.  In the standard treatment of quantum fields on
cosmological space-times, one typically works with conformal or
proper time, makes a heavy use of the causal structure made
available by the fixed background space-time, and discusses
dynamics as an unitary evolution in the chosen time variable. In
quantum geometry state of loop quantum cosmology, none of these
structures are available. Even in the `deparameterized picture'
it is a \emph{scalar field} that plays the role of internal
time; proper and conformal times are at best operators. Even
when the quantum state is sharply peaked on an effective
solution, we have only a probability distribution for various
space-time geometries; we do not have a single, well-defined,
classical causal structure. Finally, in loop quantum gravity,
dynamics is teased out of the constraint while in quantum field
theory in curved space-times it is dictated by a Hamiltonian.
These obstacles seem formidable at first, Ashtekar, Kaminski and
Lewandowski have shown that they can be overcome if one works
with spatially compact topology and focuses just on a finite
number of modes of the test field. The first assumption frees
one from infra-red issues which can be faced later while the
second restriction was motivated by the fact that, in the
inflationary scenario, only a finite number of modes of
perturbations are relevant to observations. It is important to
remove these restrictions and use the resulting framework to
analyze the questions on the quantum gravity origin of Hadamard
states and of the adiabatic regularization procedure routinely
used in cosmology.

\subsubsection{Applications}

\b \emph{The very early universe:} Because the initial
motivations for inflation are not as strong as they are
often portrayed to be, several prominent relativists were
put off by the idea. As a consequence, recent developments
in the inflationary paradigm have not drawn due attention
in general relativity circles. In my view, there is a
compelling case to take the paradigm seriously: it
predicted the main features of inhomogeneities in the
cosmic microwave background (CMB) which were subsequently
observed and which serve as seeds for structure formation.

Let me first explain this point in some detail. Note first that
one analyzes CMB inhomogeneities in terms of their Fourier modes
and observationally relevant wave numbers are in a finite range,
say $\Delta k$. Using this fact, we can write down the four
assumptions on which the inflationary scenario is based:

\noindent 1) Some time in its very early history, the universe
underwent a phase of accelerated expansion during which the
Hubble parameter H was nearly constant.\\
2) During this phase the universe is well-described by a FLRW
background space-time together with linear perturbations.\\
3) A few e-foldings before the longest wave length mode in the
family $\Delta k$ under consideration exited the Hubble radius,
these Fourier modes of quantum fields describing perturbations
were in the Bunch-Davis vacuum.\\
4) Soon after a mode exited the Hubble radius, its quantum
fluctuation can be regarded as a classical perturbation and
evolved via linearized Einstein's equations.

Then QFT on FLRW space-times and classical general relativity
imply the existence of tiny inhomogeneities in CMB which have
been seen by the 7 year WMAP data. Numerical simulations show
that these seeds grow to yield the large scale structure that is
observed today. Although the assumptions are by no means
compelling, the overall economy of thought it is nonetheless
impressive. In particular, in this paradigm, the origin of the
large scale structure of the universe lies just in vacuum
fluctuations! Therefore, it is of considerable interest to
attempt to provide a quantum gravity completion of this
paradigm.

The issues that are left open by this standard paradigm are of
two types: Particle Physics issues and Quantum Gravity issues.
Let me focus on the second for now:

\noindent 1) Initial singularity: The paradigm assumes classical
general relativity and theorems due to Borde, Guth and Vilenkin
then imply that space-time had an initial big bang singularity.
For reasons discussed in section \ref{s1}, this is an artifact
of using general relativity in domains where it is not
applicable. Therefore, one needs a viable treatment of the
Planck regime and the corresponding extension of the
inflationary paradigm.\\
2) Probability of inflation: In loop quantum cosmology, the
big bang is replaced by a quantum bounce. So it is natural
to introduce initial conditions there. Will a generic
homogeneous, isotropic initial state for the background,
when evolved, encounter a phase of slow roll inflation
compatible with the seven year WMAP data?\\
3) Trans-Planckian issues: In classical general relativity, if
we evolve the Fourier modes of interest back in time, they
become trans-Planckian. We need a quantum field theory on
\emph{quantum cosmological space-times} to adequately handle
them.\\
4) Observations: The question then is whether the initial
quantum state at the bounce, when evolved forward in time agrees
sufficiently with the Bunch Davis vacuum at the onset of
inflation so as not to contradict the observations. More
importantly, are there small deviations which could be observed
in future missions?

Recent work by Ashtekar, Sloan, Agullo, Nelson and Barreau,
Cailleteau, Grain and Mielczarek has made notable advances in
facing these questions \cite{asrev} but there are ample
opportunities for other research that will provide both a viable
quantum gravity completion of the inflationary paradigm and
potentially observable predictions.

Finally, even if loop quantum gravity does offer a viable
quantum gravity completion of the inflationary paradigm, open
issues related to particle physics will still remain. In
particular: What is the physical origin of the inflaton field?
Of the potential one must use to get a sufficiently long slow
roll? Is there only one inflaton or many? If many, what are
their interaction? What are the couplings that produce the known
particles as the inflaton oscillates around the minimum of the
potential at the end of inflation (the so-called `reheating')?
Therefore, it would be healthy to look also for alternatives to
inflation. Indeed, in the alternatives that have been advocated
by Barndenberger and others involve bouncing models and
therefore have similarities with the general loop quantum
cosmology paradigm. Because the expansion of the universe from
the bounce to the surface of last scattering in the post-bounce
branch is \emph{much} smaller than that in the inflationary
scenario, at the bounce, modes of direct interest to the CMB
observations now have physical frequencies \emph{much} below the
Planck scale. Therefore, the trans-Planckian issue is avoided
and quantum field theory in curved space-times should be viable
for these modes. This fact, coupled with the absence of
singularity, enables one to calculate a transfer matrix relating
modes in the pre-bounce epoch to those in the post-bounce epoch.
Under suitable assumptions, Brandenberger and others have shown
that this relation gives rise to a nearly scale invariant
spectrum of scalar and tensor modes in the post-bounce phase.
But the underlying premise in these calculations is that
perturbations originate in the distant past of the contracting
branch where the geometry is nearly flat and quantum fields
representing perturbations are taken to be in their vacuum
state. This idea that the entire evolution from the distant past
in the contracting phase to the bounce is well described by a
homogeneous model with small perturbations is not at all
realistic. But since the general paradigm has several attractive
features, it would be of considerable interest to investigate
whether loop quantum cosmology bounces allow similar
alternatives to inflation without having to assume that
non-linearities can be neglected throughout the pre-bounce
phase.\\

\b \emph{Black hole evaporation: The issue of the final state:}
Black hole thermodynamics was initially developed in the context
of stationary black holes. Indeed, until relatively recently,
there were very few analytical results on dynamical black holes
in classical general relativity. This changed with the advent of
dynamical horizons which provide the necessary analytical tools
to extract physics from numerical simulations of black hole
formation and evaporation. It also led to some new insights on
the fundamental side. in particular, it was also shown that the
first law can be extended to these time-dependent situations and
the leading term in the expression of the entropy is again given
by $\textstyle{a_{\rm hor}/4\lp^2}$ \cite{akrev}. Hawking
radiation will cause the horizon of a large black hole to shrink
\emph{very} slowly, whence it is reasonable to expect that the
description of the quantum horizon geometry can be extended from
isolated to dynamical horizons in this phase of the evaporation.
The natural question then is: Can one describe in detail the
black hole evaporation process and shed light on the issue of
information loss?

\begin{figure}[]
  \begin{center}
    \includegraphics[width=1.5in,angle=0]{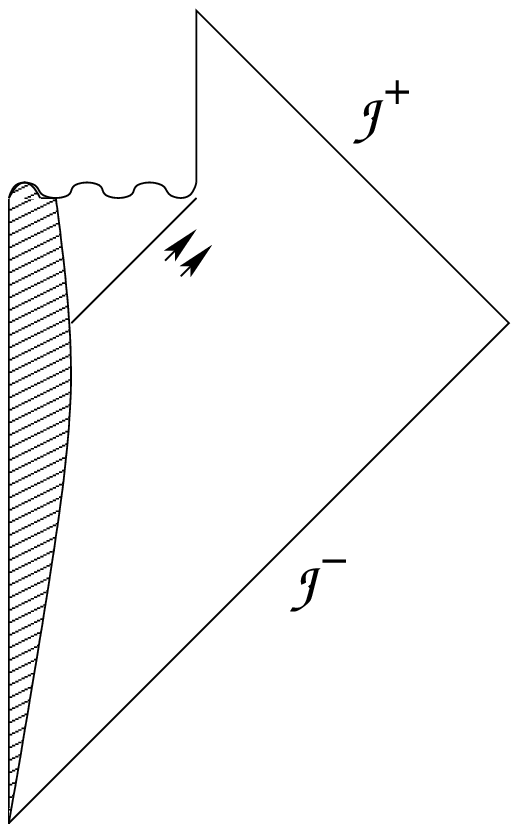} \hspace{3cm}
    \includegraphics[width=1.5in,angle=0]{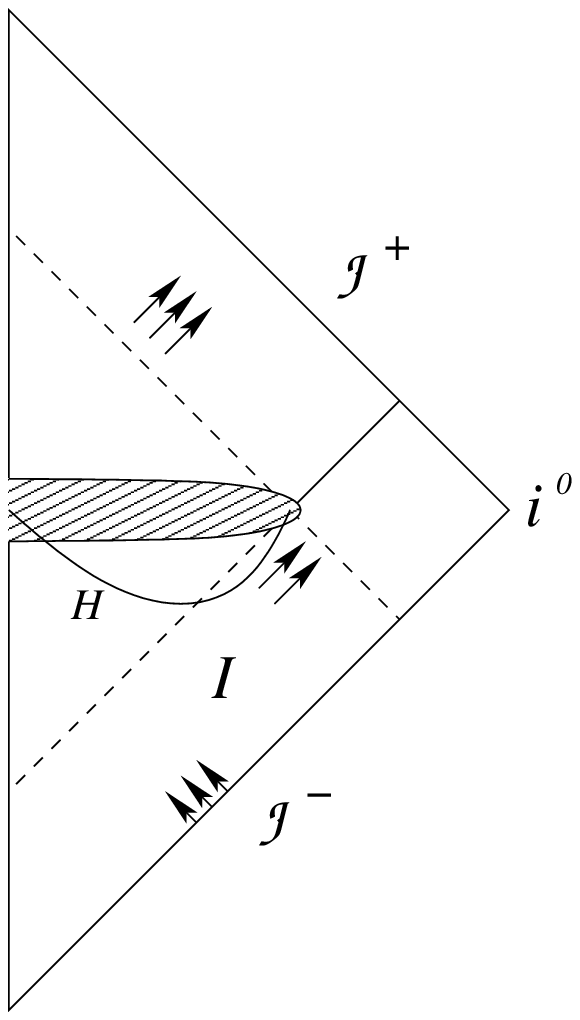}
\caption{Conjectured space-time diagrams of evaporating black
holes in full quantum theory.\\ (a) Left figure: Standard paradigm,
originally proposed by Hawking. Information is lost because
part of the incoming state
on $\mathcal{I}^-$  falls into the part of the future singularity that
is assumed to persist in the full quantum gravity theory.\,\,
(b) Right figure: New paradigm motivated by the singularity
resolution in LQC. What forms and evaporates is a dynamical horizon $H$.
Quantum space-time is larger and the incoming information on the full
$\mathcal{I}^-$ is adequately recovered on the $\mathcal{I}^-$ of this
larger space-time.}
\label{Traditional}
 \end{center}
\end{figure}

The space-time diagram of the evaporating black hole,
conjectured by Hawking, is shown in the left-hand drawing in
Fig. \ref{Traditional}. It is based on two ingredients: i)
Hawking's original calculation of black hole radiance, in the
framework of quantum field theory on a \emph{fixed} background
space-time; and ii) heuristics of back-reaction effects which
suggest that the radius of the event horizon must shrink to
zero. It is generally argued that the semi-classical process
depicted in this figure should be reliable until the very late
stages of evaporation when the black hole has shrunk to Planck
size and quantum gravity effects become important. Since it
takes a very long time for a large black hole to shrink to this
size, one then argues that the quantum gravity effects during
the last stages of evaporation will not be sufficient to restore
the correlations that have been lost due to thermal radiation
over such a long period. Thus there is loss of information.
Intuitively, the lost information is `absorbed' by the
`left-over piece' of the final singularity which serves as a new
boundary to space-time.

However, loop quantum gravity considerations suggest that this
argument is incorrect in two respects. First, the semi-classical
picture breaks down not just at the end point of evaporation but
in fact \emph{all along what is depicted as the final
singularity}. Using ideas from quantum cosmology, the interior
of the Schwarzschild horizon was analyzed in the context of loop
quantum gravity by Ashtekar, Bojowald, Modesto, Vandersloot and
others. This analysis is not as complete or refined as that in
the cosmological context. But the qualitative conclusion that
the singularity is resolved due to quantum geometry effects is
likely to be robust. If so, the space-time does \emph{not} have
a singularity as its final boundary. The second limitation of
this semi-classical picture is its depiction of the event
horizon. The notion of an event horizon is teleological and
refers to the \emph{global} structure of space-time. Resolution
of the singularity introduces a domain in which there is no
classical space-time, whence the notion ceases to be meaningful;
it is simply `transcended' in quantum theory. Using these
considerations Ashtekar and Bojowald introduced a new paradigm
for black hole evaporation in loop quantum gravity, depicted in
the right hand drawing of Fig. \ref{Traditional}: Now, it is the
dynamical horizon that evaporates with emission of quantum
radiation, and the initial pure state evolves to a final pure
state on the future null infinity of the extended space-time.
Thus, there is no information loss. In this paradigm, the
semi-classical considerations would not simply dismissed; they
would be valid in certain space-time regions and under certain
approximations. But for fundamental conceptual issues, they
would not be inadequate.

However, this is still only a paradigm and the main challenge is
to develop it into a detailed theory. Just as Wheeler's `It from
Bit' ideas were transformed into a detailed theory of quantum
horizon geometry, it should be possible to construct a detailed
theory of black hole evaporation based on this paradigm. More
recently, this paradigm was put on a firm footing by Ashtekar,
Taveras and Varadarajan in the case of 2-dimensional black holes
first introduced by Callen, Giddings, Harvey and Strominger. The
model is interesting especially because its action and equations
of motion closely mimic those governing 4-dimensional,
spherically symmetric black holes formed by the gravitational
collapse of a scalar field. Ashtekar, Pretorius and Ramazanoglu
have used a combination of analytical and numerical methods to
analyze the mean field approximation in complete detail. It
explicitly shows that some of the common assumptions regarding
effects of including back reaction, discussed in the last
paragraph, are incorrect. This analysis further reinforces the
paradigm of the figure on the right. It is therefore of
considerable interest to extend all this analysis to four
dimensions in the loop quantum gravity setting. The very
considerable work on spherically symmetric midi-superspaces by
Gambini, Pullin, Bojowald and others will serve as a point of
departure for this analysis.\\

\b \emph{Contact with low energy physics:} Spin foam models
provide a convenient arena to discuss issues such as the
graviton propagator, $n$ point functions and scattering, that
lie at the heart of perturbative treatments. At first, it seems
impossible to have non-trivial $n$-point functions in a
diffeomorphism invariant theory. Indeed, how could one say that
the 2-point function falls off as $1/r^n$ when the distance $r$
between the two points has no diffeomorphism invariant meaning?
Thanks to a careful conceptual set-up by Oeckl, Colosi, Rovelli
and others, this issue has been satisfactorily resolved. To
speak of $n$-point functions, one needs to introduce a boundary
state (in which the expectation values are taken) and the notion
of distance $r$ descends from the boundary state. Interestingly,
a detailed calculation of the 2-point function brought out some
limitations of the Barrett-Crane model and provided new impetus
for the EPRL and FK models. As Perini's talks at this school
showed, these calculations by Bianchi, Ding, Magliaro and Perini
strongly indicate that, to the leading order, a graviton
propagator with the correct functional form and tensorial
structure will arise from these models.

However, these calculations can be improved in a number of
respects and their full implications have yet to be properly
digested. In particular, one needs a better handle on
contributions from 2-complexes with large numbers of vertices
and the physics of the sub-leading terms. These terms seem to be
sensitive to the choice of the boundary state and there isn't a
canonical one representing Minkowski space. Therefore,
comparison with the standard perturbation theory in Minkowski
space is difficult. This is a fertile and important area for
further research. Indeed the key challenge in this area is to
`explain' why perturbative quantum general relativity fails if
the theory exists non-perturbatively. As mentioned in section
\ref{s1}, heuristically the failure can be traced back to the
insistence that the continuum space-time geometry is a good
approximation even below the Planck scale. But a more detailed
answer is needed. For example, is it because, as developments in
the asymptotically safe scenarios indicate
\cite{reuterrev,reuter}, the renormalization group has a
\emph{non-Gaussian} fixed point?\\

\b  \emph{Unification.}  Finally, there is the issue of
unification. At a kinematical level, there is already an
unification because the quantum configuration space of general
relativity is the same as in gauge theories which govern the
strong and electro-weak interactions. But the non-trivial issue
is that of dynamics. To conclude, let us consider a speculation.
One possibility is to use the `emergent phenomena' scenario
where new degrees of freedom or particles, which were not
present in the initial Lagrangian, emerge when one considers
excitations of a non-trivial vacuum. For example, one can begin
with solids and arrive at phonons; start with superfluids and
find rotons; consider superconductors and discover cooper pairs.
In loop quantum gravity, the micro-state representing Minkowski
space-time will have a highly non-trivial Planck-scale
structure. The basic entities will be 1-dimensional and
polymer-like. one can argue that, even in absence of a detailed
theory, the fluctuations of these 1-dimensional entities should
correspond not only to gravitons but also to other particles,
including a spin-1 particle, a scalar and an anti-symmetric
tensor. These `emergent states' are likely to play an important
role in Minkowskian physics derived from loop quantum gravity. A
detailed study of these excitations may well lead to interesting
dynamics that includes not only gravity but also a select family
of non-gravitational fields. It may also serve as a bridge
between loop quantum gravity and string theory. For, string
theory has two a priori elements: unexcited strings which carry
no quantum numbers and a background space-time. Loop quantum
gravity suggests that both could arise from the quantum state of
geometry, peaked at Minkowski (or, de Sitter) space. The
polymer-like quantum threads which must be woven to create the
classical ground state geometries could be interpreted as
unexcited strings. Excitations of these strings, in turn, may
provide interesting matter couplings for loop quantum gravity.

\subsubsection{Some Final Remarks}

{}From examples discussed in this section it is clear that loop
quantum gravity has witnessed significant advances over the last
decade, both in its foundations and applications. It is
therefore important for the community to make sustained progress
on directions that have already been opened up. Of course one
constantly needs an influx of new ideas. But it would be a
mistake if a significant fraction of the community focuses on
constructing new models every few months, making a first stab
and then passing on to the next model. The cumulative results in
the main stream development of loop quantum gravity now carry
sufficient weight for us to take the basic ideas seriously and
continue to develop them by attacking the hard conceptual and
technical open issues. Examples of such issues are: Finding
principles and strategies to significantly narrow the
ambiguities in the definition of the Hamiltonian constraint;
exploring the role of supersymmetry; sharpening the set of
quantum geometries to sum over, and addressing the problem of
convergence in spin foam models; analyzing the renormalization
group flows in group field theory; understanding the dependence
of the n-point functions on the choice of the boundary state;
developing approximation methods to calculate S-matrix from spin
foams and pin-pointing why the standard perturbative treatments
fail; fully incorporating matter fields in spin foams,
particularly scalar fields; constructing effective field
theories to adequately describe low energy physics; finding the
detailed relation between loop quantum gravity and loop quantum
cosmology; constructing a detailed completion of the
inflationary paradigm in the Planck regime; exploring its
observable consequences in the very early universe; ... The list
is long enough to keep young researchers busy and happy for
quite a while! Furthermore, in this work, it is important to
keep focus on physical issues and try to solve problems of
\emph{direct physical interest}. Developing formalism is
important because it streamlines the ideas and procedures. But
it is not an end in itself. Indeed, it is of little use unless
it leads to answers to the long standing physical questions.

Finally, there is a complementary direction. Because the
mathematics underlying loop quantum gravity is rigorous, the
subject is now begun to make inroads into other areas of
mathematical physics and mathematics itself. For example, there
is now literature on spin networks, quantum topology and
computing, spin foams and non commuting geometry, Holst action
and asymptotic safety, loop quantum gravity and topos theory,
spectral triples over the space of generalized connections etc.
from researchers like Kauffmann, Marcolli, Reuter, Dahlen,
Aastrup and Grimstrup from outside the traditional loop quantum
gravity community. Conversely, some loop quantum gravity
researchers such as Ma, Gambini, Pullin, Singh, Dittrich,
Freidel and Fleischhack are applying techniques developed in the
field to other areas such as the $f(R)$ theories that rose to
prominence in cosmology, generalized quantum mechanics,
statistical mechanics, non-commutative geometry, gauge theories
and geometry. Thus there is ample evidence that the subject is
now sufficiently mature to have applications to other areas. In
these explorations, it is important to focus on problems that
other communities consider as important in their areas. In my
view, this `outward bound' spirit is the second pillar on which
further development of the field will rest.

\section*{Acknowledgements:}

My understanding of classical and quantum gravity has deepened
through discussions with a large number of colleagues; many of
whom were at the Zakopane School. Parts of this overview are
updated versions of some of the material in author's article
\cite{njp}. I would like to thank Miguel Campiglia, Kristina
Giesel, Alok Laddha and especially Carlo Rovelli for their
comments on the manuscript. This work was supported in part by
the European Science Foundation through its network
\emph{Quantum Geometry and Quantum Gravity}, NSF grant PHY
0090091 and the Eberly research funds of The Pennsylvania State
University.

\end{document}